\documentclass{iau} 
\usepackage{graphicx}

\title[Open star clusters and Galactic structure] 
{Open star clusters and Galactic structure}

\author[Yogesh C. Joshi]   
{Yogesh C. Joshi$^1$}

\affiliation{$^1$Aryabhatta Research Institute of Observational Sciences, Nainital, India - 263002 \\email: {\tt yogesh@aries.res.in}}

\pubyear{2017}
\volume{330}  
\jname{Astronomy and Astrophysics in the GAIA sky}
\editors{A. Recio-Blanco, P. de Laverny, A. Brown and, T. Prusti, eds.}
\begin{document}

\maketitle

\begin{abstract}
In order to understand the Galactic structure, we perform a statistical analysis of the distribution of various cluster parameters based on an almost complete sample of Galactic open clusters yet available. The geometrical and physical characteristics of a large number of open clusters given in the MWSC catalogue are used to study the spatial distribution of clusters in the Galaxy and determine the scale height, solar offset, local mass density and distribution of reddening material in the solar neighbourhood. We also explored the mass-radius and mass-age relations in the Galactic open star clusters. We find that the estimated parameters of the Galactic disk are largely influenced by the choice of cluster sample.
\end{abstract}

\firstsection 
%
\section{Introduction}
Open clusters are distributed throughout the Galactic disk and span a wide range in ages hence they are useful tool to study the effects of dynamical evolution of the Galactic disk (Carraro et al. 1998). Even though much progress has been made in understanding the general properties of open star clusters and Galactic structure but continuous increase of the cluster sample and refinement of cluster parameters pave way for a better understanding of the Galactic structure (Joshi et al. 2016). The observed distribution of cluster parameters and correlations between various parameters offer important empirical constraints not only on the cluster formation, but also on the history of the Galaxy. However, any meaningful study of star clusters is very much dependent on the accurate determination of the cluster parameters as well as the homogeneity and degree of completeness of the cluster sample. 
\section{Data}
To understand the Galactic structure, statistical analysis of a large number of star clusters in the Galaxy is very important because individual cluster parameters may be affected by the considerable uncertainty in their determination. We used MWSC catalogue that gives the physical parameters of 3208 star clusters (Kharchenko et al. 2013, Schmeja et al. 2014, Scholz et al. 2015) which were determined in homogeneous manner since parameters of all the clusters are estimated through the same technique. Our analysis found that MWSC catalogue is complete only up to a distance of 1.8 kpc which is similar to Dias catalogue\footnote{http://www.astro.iag.usp.br/ocdb} on open clusters (Joshi 2005, 2007). After rejecting the dubious clusters in the MWSC catalogue within the completeness limit of 1.8 kpc, we used 1218 open clusters to study the Galactic structure.
\section{Results and Conclusion}
Although a detailed study to understand the Galactic structure in the solar neighbourhood is reported in Joshi et al. (2016) based on MWSC survey catalogue, a brief summary of the results is given below. 
\begin{enumerate}
\item Assuming a uniform density model for the distribution of clusters, the catalogue of open clusters is found to be complete only up to 1.8 kpc distance.
\item We found a huge deficiency of old open clusters ($>$ 700 Myr) within 500 pc from the Sun, having only 6 out of a total 91 clusters (6.6\%) are located within this distance. It is understood that the most of the massive clusters in the solar neighbourhood would dissolve even before reaching the age of 1 Gyr causing the deficiency of old clusters in the solar neighbourhood.
\item The distribution of clusters in various longitude bins revealed that the maximum clusters are located in the region of around $125^{o}$ (Cassiopeia), 210 deg (Monoceros), 240 deg (Canis Major), and 285 deg (Carina) while deep minima is primarily found in two longitudes, around 50 deg (Sagitta) and 150 deg (Perseus).
\item The maximum Galactic absorption is seen towards $l\sim40^o$ and minimum towards $l\sim220^o$.
\item The solar offset is found to be in the range of 6 to 15 kpc, however, it strongly depends on the choice of stellar population, data sampling and estimation method.
\item The disk scale height is found to be $z_h = 64\pm2$ pc, however, it decreases to $z_h=60\pm2$ pc when we only consider clusters younger than 700 Myr. Although it is close to generally accepted value of $z_h$ deduced from the cluster sample but it is noticed that the difference in measurement of $z_h$ is strongly influenced by the selection criteria of the cluster sample.
\item It is found that scale height increases with mean age of the clusters and, in general, $z_h$ increases from $\sim$40 pc at 1 Myr to $\sim$75 pc at 1 Gyr. Furthermore, $z_h$ is found to be larger in the direction of Galactic anti-center than the Galactic center and, on an average, $z_h$ is more than twice as large as in the outer region than in the inner region of the solar circle, except for the youngest population of clusters.
\item We estimated a local mass density of $\rho_0$~=~0.090~$\pm$~0.005~$M_{\odot}/{\rm pc}^3$ but did not find any significant contribution of dark matter in the solar neighborhood.
\item The reddening in the direction of clusters suggests a strong correlation with their vertical distance from the Galactic plane with a respective slope of $dE(B-V)/dz$ = 0.40$\pm$0.04 and 0.42$\pm$0.05 mag/kpc below and above the Galactic Plane. 
\item We observed a linear mass-radius and mass-age relations and derived slopes of $\frac{dR}{d(logM)}=2.08\pm0.10$ and $\frac{d(logM)}{d(logT)}=0.36\pm0.05$ for the Galactic open clusters.
\end{enumerate}

\section*{Acknowledgments}
I wish to thank IAU for providing financial support to attend the conference.

\end{document}